\documentclass[aps,prb,twocolumn,superscriptaddress,showpacs]{revtex4}

\usepackage{graphicx}
\usepackage{epsfig}
\usepackage{natbib}

\begin{document}


\title{Entanglement between charge qubits induced by a common dissipative environment}

\author{L.D. Contreras-Pulido}
\email[(present) ]{  debora.contreras@icmm.csic.es}
\affiliation{Departamento de Teor\'ia de la Materia Condensada,
Instituto de Ciencia de Materiales de Madrid, CSIC, Cantoblanco
28049, Madrid, Spain} \affiliation{Centro de Investigaci\'on
Cient\'ifica y de Educaci\'on Superior de Ensenada, Apartado Postal
2732, Ensenada, B.C. 22860, M\'exico}

\author{R. Aguado}
\email[]{raguado@icmm.csic.es}
\affiliation{Departamento de Teor\'ia de la Materia Condensada,
Instituto de Ciencia de Materiales de Madrid, CSIC, Cantoblanco
28049, Madrid, Spain}


\date{\today}

\begin{abstract}
We study entanglement generation between two charge qubits due to
the strong coupling with a common bosonic environment (Ohmic bath).
The coupling to the boson bath is a source of both quantum noise
(leading to decoherence) and an indirect interaction between qubits.
As a result, two effects compete as a function of the coupling
strength with the bath: entanglement generation and charge
localization induced by the bath. These two competing effects lead
to a non-monotonic behavior of the concurrence as a function of the
coupling strength with the bath. As an application, we present
results for charge qubits based on double quantum dots.
\end{abstract}


\maketitle

\section{INTRODUCTION}
Solid state nanostructures have become promising candidates for
quantum information processing \cite{ref1}, with basic operations
like single-qubit manipulation and readout having been demonstrated
during the last few years. However, to go beyond single-qubit
manipulations, and study effects such as entanglement generation and
quantum gate operations, one needs some kind of interaction between
the qubits. Although this interaction usually comes from a direct
coupling between qubits (like the Coulomb interaction for charge
qubits or exchange coupling for spin qubits), entanglement can be
also generated by coupling two qubits (which do not interact with
each other) to a common third system
\cite{plenio99,milburn,nicolosi,chinos,kraus,paternostro3,plenio2002,zheng2000,kim2002,oh}.
In most of these studies, the indirect interaction comes from the
coupling to one or a few external degrees of freedom. Examples
include the coupling to electromagnetic modes in a cavity (see, for
example, Ref. \onlinecite{plenio99}, where the authors study
entanglement of atoms within a single-mode cavity field) or to a
harmonic oscillator representing a mode in a thermal environment
\cite{kim2002,oh}. Importantly, entanglement can be also induced
when the environment is made by an infinitely large number of
degrees of freedom, namely a bath, as demonstrated by Braun in Ref.
\onlinecite{braun}. This is an important case because entanglement
is generated exclusively by \emph{incoherent means}. In this
context, different works have studied the coupling of two
non-interacting qubits to fermionic
\cite{craig,mozy,piermarocchi,privman98,lambert} or bosonic
\cite{braun,solenov06,solenov07,vorrath2003,oh} baths.

Indirect qubit interactions have attracted attention recently
because the information distribution among distant entangled
particles is the base of quantum cryptography \cite{ekert}, quantum
teleportation \cite{bennett93,bouwmesteer1,nielsen} quantum dense
code \cite{bennett92,mattle,nielsen}, different processes proposed
for testing Bell inequalities \cite{bose,braunstein,chsh,gisin} and
even certain steps within quantum computation algorithms. The
possibility of entangling two quantum systems which do not interact
directly is therefore highly desirable, with various aspects of
current interest like "entanglement swapping"
\cite{bose,lee,pan,zou,zukow} and "entanglement transfer"
\cite{paternostro1,paternostro2}.

In this paper we study entanglement generation between two charge
qubits due to the strong coupling with a common bosonic environment
(Ohmic bath). For concreteness, we focus on charge qubits based on
double quantum dots (DQDs) but we point out, that our results can
also be applied to Cooper Pair Boxes in a resistive environment. In
a DQD, the electron charge degree of freedom is used to construct a
qubit
\cite{vanderWiel,fujisawa2004a,experimentos,zanardi,tanamoto2000,brandesrep},
with logical states $|0\rangle$ and $|1\rangle$ corresponding to the
localization of one excess electron on each one of the quantum dots
(QD). One of the advantages of these charge qubits is their
controllability through external voltages handling, as demonstrated
in recent experiments \cite{experimentos} where the charge has been
coherently manipulated.
\begin{figure}
  \begin{center}
    \epsfig{file=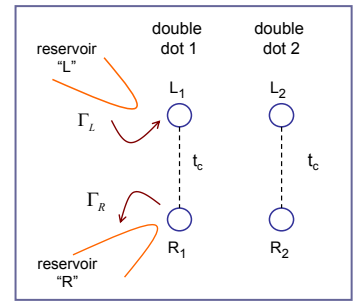,height=5cm}
    \caption{Two independent double quantum dots coupled to a common phonon environment. Interdot hopping, $t_c$, is allowed only in each double
    dot. The first double quantum dot is open to electron reservoirs, with probabilities for tunneling in and out given by $\Gamma_L$ and
$\Gamma_R$, respectively.
    }
    \label{fig1}
\end{center}
\end{figure}
We model the two DQD system as two independent two-level systems
strongly coupled to the same Ohmic bath (two spin-boson models). In
addition, we consider that one of the DQDs is coupled to electron
reservoirs \cite{ramon04} (Fig. \ref{fig1}), in order to allow
electronic transport. The coupling to electronic reservoirs is
treated using a Markovian approach
\cite{brandesrep,ramon04,brandes99,stoofynazarov}, which is valid in
the sequential tunneling limit and large bias voltages. The
non-Markovian character of the strong coupling with the boson bath
is, on the other hand, taken into account by using a polaron
approach \cite{brandesrep,ramon04,brandes99,platero,mahan}. As a
result of this strong coupling, an indirect Ising-like interaction
between qubits is induced by the bath.

By combining both Markovian and non-Markovian approximations, we
derive a master equation for the reduced density matrix of the
system, including boson correlation functions in Laplace space. The
resulting density matrix is used to calculate the degree of
entanglement (quantified by Wotters' concurrence \cite{wootters98})
as well as the probability for each one of the Bell states as a
function of the coupling strength by the bath.

Our results complement previous work by Vorrath and Brandes in Ref.
\onlinecite{vorrath2003} who studied a similar problem within a
Markovian approach. We also mention some recent works
\cite{solenov06,solenov07,oh} in which related models are treated.

The paper is organized as follows: in section II the model
describing the DQD coupled to both, electronic reservoirs and the
bosonic environment is discussed. We also present in Section II the
general solution scheme for the density matrix equations. The
coupling with the leads is treated by using a standard Born-Markov
approximation whereas the strong coupling with
the bath is treated within a polaron approach. 
Section III shows the main
results obtained, and finally we conclude in section IV.

\section{MODEL}
An array of two parallel DQDs in the strong Coulomb Blockade regime,
and coupled to the same bosonic environment, is considered. Interdot
tunneling is allowed only in each double dot, defining an array of
two charge qubits (Fig. \ref{fig1}). The first DQD is weakly coupled
to two electron reservoirs in such way that electronic transport
through this double dot is possible (the excess charge in this DQD
fluctuates between one and zero). The second DQD is closed and,
therefore, has always one excess electron. Note that such
configuration is close to the one realized in very recent
experiments \cite{fujisawaapl}.The Hilbert space includes
two-particles states $|1\rangle=|L_{1}L_{2}\rangle$,
$|2\rangle=|L_{1}R_{2}\rangle$, $|3\rangle=|R_{1}L_{2}\rangle$ and
$|4\rangle=|R_{1}R_{2}\rangle$ (where $L_{i} (R_{i})$ represents the
charge localized in the upper (lower) QD of the $i$-th DQD), as well
as one particle states $|5\rangle=|0_{1}L_{2}\rangle$ and
$|6\rangle=|0_{1}R_{2}\rangle$ (where $0$ means no extra electron in
the first DQD). The completeness of the system is therefore given by
$\sum_{k=1}^{6}|k\rangle\langle k|$.

The total Hamiltonian describing this system reads
\begin{equation}
H=H_{sys}+H_{res}+H_B+H_{SR}+H_{SB} \label{eq6}.
\end{equation}

The free part of the Hamiltonian, i.e without couplings, contains
three terms.  The first term corresponds to the Hamiltonian of two
independent DQDs, which in pseudo-spin language can be written as

\begin{equation}
H_{sys}=\displaystyle
\sum_{i}^{2}\frac{1}{2}\Delta\varepsilon_i\sigma_z^i+t_c\sigma_x^i,
\label{eq1}
\end{equation}
where $\Delta\varepsilon_i$ is the energy difference between quantum
dots of each pair being
$\Delta\varepsilon_1=\varepsilon_{L_1}-\varepsilon_{R_1}$ and
$\Delta\varepsilon_2=\varepsilon_{L_2}-\varepsilon_{R_2}$
$(\varepsilon_{i_j}$ is the on-site energy of the \emph{i}-th QD of
the pair $j$), $\sigma_j^i$ is the \emph{j}-th Pauli matrix acting
on each DQD, $t_c$ is the electron tunneling amplitude which is
considered identical for both DQDs \cite{ramon04,brandes99}.

The Hamiltonian of the reservoirs, referred as "L" and "R", reads
\cite{brandesrep,ramon04,brandes99}
\begin{equation}
H_{res}=\displaystyle
\sum_{k}\{\epsilon_k^Lc_{k,L}^{\dagger}c_{k,L}+\epsilon_k^Rc_{k,R}^{\dag}c_{k,R}\},
\label{eq2}
\end{equation}
where $c_{k,\beta}^{\dag}$ and $c_{k,\beta}$ are fermion creation
and annihilation operators in lead $\beta$ with corresponding energy
$\epsilon_k^\beta$. Finally, the third term corresponds to the boson
bath, which is described as a set of harmonic oscillators with
frequency $\omega_q$:

\begin{equation}
H_B=\displaystyle \sum_{q}\hbar\omega_{q}a_q^{\dag}a_q
\label{eq4}
\end{equation}
$a_q^{\dag} (a_q)$ is the annihilation (creation) boson operator
\cite{ramon04,weiss}.

As we have mentioned already, we take into account the coupling of
the system to both electronics reservoirs and a boson bath. The
first coupling is given by
\begin{equation}
H_{SR}=\displaystyle \sum_{k,i\in 1,2}\{V_k^L(c_{k,L}^{\dag}s_{L,i}+
c.c)+V_k^R(c_{k,R}^{\dag}s_{R,i}+c.c.)\}, \label{eq3}
\end{equation}
with $V_k^\beta$ being the coupling with the lead $\beta$. The
Lindblad-type operators $s_{L,i}$ ($s_{R,i}$) describe tunneling
into (out of) the first DQD taking into account the two possible
configurations ($\sum_{i\in 1,2}$) in the second DQD, namely
$s_{L,1}=|5\rangle\langle1|$, $s_{L,2}=|6\rangle\langle2|$,
$s_{R,1}=|5\rangle\langle3|$ and $s_{R,2}=|6\rangle\langle4|$.
Although we consider the coupling of only one of the DQDs to
reservoirs, the generalization for both double dots is
straightforward.

The electron-boson interaction is described with a spin-boson
Hamiltonian, where the bath "force" operator
$\xi^{i}=\sum_{q}\gamma_{q}^{i}(a_q^{\dag}+a_q)$ couples linearly to
each qubit' s $\sigma_z^{i}$ \cite{weiss,legget} (here we consider
that such interaction is identical for both DQDs,\cite{vorrath05}
$\gamma_{q}^{i}=\gamma_{q}$):

\begin{equation}
H_{SB}=\displaystyle
\sum_{q}\sum_{i}\frac{1}{2}\gamma_{q}\sigma_{z}^{i}(a_q^{\dag}+a_q),
\label{eq5}
\end{equation}
Note that this coupling which is longitudinal in the local basis of
each qubit, contains both longitudinal and traversal components in
the basis that diagonalizes the qubit Hamiltonian \cite{gschon}. The
bath effects can be encapsulated in the spectral density
$J(\omega)=\sum_q\gamma_{q}^{2}\delta(\omega-\omega_q)$. In the
following, we use a generic ohmic bath: $J(\omega)=2\alpha \omega
e^{-\omega / \omega_c}$, where $\omega_c$ is a cutoff frequency and
$\alpha$ is a dimensionless parameter which reflects the dissipation
strength \cite{weiss,legget,thorwart04a}. As we shall see in the
next subsection, the coupling of both qubits to the same quantum
heat bath leads to decoherence \emph{and} to an effective
interaction between qubits.

\subsection{Polaron Transformation}
Due to the strong coupling of the qubits with the boson bath, a
proper description of the system must take into account
non-Markovian effects. Among the different approaches available to
deal with this \cite{nonmarkov1,nonmarkov2,nonmarkov3,breuer}, we
use a "polaron transformation" \cite{mahan}, which for an arbitrary
operator $O$ is given by:
\begin{eqnarray}
\overline{O}=&&e^{S}Oe^{-S}\;,
\nonumber  \\
S=&&\sum_{q}\sum_{i}\frac{1}{2}\sigma_{z}^{i}\frac{\gamma_{q}}{\omega_q}(a_q^{\dag}-a_q).
\label{eq7}
\end{eqnarray}
This approach, which is well-known for treating problems in which
bosonic modes couple to localized electronic states, has been
successfully used for studying single \cite{wingreen,mitra} and
double quantum dots\cite{ramon04,brandesrep,brandes99,platero}
strongly coupled to a bath of phonons.

Applying the transformation in Eq. (\ref{eq7}) to the relevant
operators in our model, we obtain the transformed operators:

\begin{eqnarray}
\overline{\sigma}_{z}^{\thinspace{i}}=&&\sigma_{z}^{i}
\nonumber  \\
\overline{\sigma}_{x}^{\thinspace{i}}=&&\sigma_{+}^{i}X+\sigma_{-}^{i}X^{\dagger}
\nonumber \\
\overline{a}_q=&&a_q-\frac{1}{2}\frac{\lambda_q}{\omega_q}\sum_i\sigma_{z}^{i}\\
\label{eqop} \overline{s}_{L,i}=&&s_{L,i}e^{-A/2}
\nonumber  \\
\overline{s}_{R,i}=&&s_{R,i}e^{A/2} \nonumber
\end{eqnarray}
where $\sigma_{+}^{i}$ and $\sigma_{-}^{i}$ are the ladder spin
operators on each DQD. The operators $X$ and $X^{\dagger}$ are
polaronic phases\cite{mahan} given by
\begin{equation}
X=e^A, \label{eqpolaron}
\end{equation}
with
\begin{eqnarray}
A=\sum_q\gamma_q\left(a_q^{\dagger}-a_q\right).\nonumber\\
\end{eqnarray}
By substituting these \emph{transformed} operators into the
equations (\ref{eq1}) to (\ref{eq5}), we obtain the effective
Hamiltonian:

\begin{eqnarray}
\overline{H}=&&\overline{H}_0+\overline{H}_T+\overline{H}_{SR}
\label{eq8}\\
\overline{H}_0=&&\sum_{i}\frac{1}{2}\Delta\varepsilon_i\sigma_z^i-\frac{1}{4}\kappa
\sum_{i,j}\sigma_z^i\sigma_z^j+H_B+H_{res} \label{eq9}\\
\overline{H}_T=&&\sum_{i}t_c(\sigma_+^{i}X+\sigma_-^{i}X^{\dag})
\label{eq10}
\end{eqnarray}

The effect of the canonical transformation is threefold:

(i) The electron-boson interaction $H_{SB}$ has been transformed
away.

(ii) The state of the bosonic system is strongly modified every time
an electron tunnels between dots (boson "shake-up")\cite{footnote}.
As a result, the interdot tunneling amplitude (Eq. \ref{eq10})
becomes renormalized with environment-dependent phases through the
operators $X=e^A$. These time-dependent exponential phases, which
appear as a result of the non-perturbative treatment of the
electron-boson interaction, lead to non-trivial effects. In
particular, this implies that non-Markovian effects become relevant
and need to be considered in the dynamics of the reduced density
matrix. Note also that, in principle, this renormalization of
tunneling has to be taken into account also in $\overline{H}_{SR}$
through the operators $\overline{s}_{L,i}= s_{L,i}e^{-A/2}$ and
$\overline{s}_{R,i}=s_{R,i}e^{A/2}$. However, this is no longer true
in the limit of large bias voltages, where the coupling to the
reservoirs becomes Markovian (see the next subsection).

(iii) The transformed Hamiltonian contains an \emph{effective
interaction} between qubits $H_{eff}=-\frac{\kappa}{4}
\sum_{i,j}\sigma_z^i\sigma_z^j$ due to the coupling with the common
bath; this interaction has an Ising form and depends on the
parameter $\kappa=\sum_{q}\gamma_{q}^{2}/\omega_q$ (which for Ohmic
dissipation used here reads $\kappa=2\alpha\omega_c$) favoring
states with the same charge distribution in both DQDs.

\subsection{Master equation}

We define the total density operator of the open system as
$\chi(t)=e^{-iHt}\chi(0)e^{iHt}$ which, after the transformation in
Eq. (\ref{eq7}), can be written in the interaction picture as
\cite{brandesrep,brandes99}
$\widetilde{\chi}=e^{i\overline{H}_{0}t}\overline{\chi}(t)e^{-i\overline{H}_{0}t}$,
with
$\overline{\chi}(t)=e^{-i\overline{H}t}\overline{\chi}(0)e^{i\overline{H}t}$.
By taking the partial trace over the reservoir degrees of freedom,
the reduced density matrix (RDM) of the two DQDs plus the boson bath
is obtained as $\widetilde{\rho}(t)=Tr_{res}{\widetilde{\chi}(t)}$.
Applying the second order Born approximation we obtain the equation
of motion for $\widetilde{\rho}(t)$ as:

\begin{widetext}
\begin{eqnarray}
\frac{d}{dt}\widetilde{\rho}(t)=&&-i[\widetilde{H}_T(t),\widetilde{\rho}(t)]\nonumber\\
&&-\sum_{k,i\in 1,2,j\in
L,R}\int_{0}^{t}dt'\thinspace|V_{k}^{j}|^{2}f^{j}(\epsilon_{k}^{j})e^{i\epsilon_{k}^{j}(t-t')}\{\widetilde{s}_{j,i}(t)\widetilde{s}_{j,i}^{\dag}(t')\widetilde{\rho}(t')-\widetilde{s}_{j,i}^{\dag}(t')\widetilde{\rho}(t')\widetilde{s}_{j,i}(t)\}
\nonumber
\\
&&-\sum_{k,i\in 1,2,j\in
L,R}\int_{0}^{t}dt'\thinspace|V_{k}^{j}|^{2}[1-f^{j}(\epsilon_{k}^{j})]e^{-i\epsilon_{k}^{j}(t-t')}\{\widetilde{s}_{j,i}^{\dag}(t)\widetilde{s}_{j,i}(t')\widetilde{\rho}(t')-\widetilde{s}_{j,i}^{\dag}(t')\widetilde{\rho}(t')\widetilde{s}_{j,i}(t)\}
\nonumber
\\
&&-\sum_{k,i\in 1,2,j\in
L,R}\int_{0}^{t}dt'\thinspace|V_{k}^{j}|^{2}f^{j}(\epsilon_{k}^{j})e^{i\epsilon_{k}^{j}(t-t')}\{\widetilde{\rho}(t')\widetilde{s}_{j,i}(t')\widetilde{s}_{j,i}^{\dag}(t)-\widetilde{s}_{j,i}^{\dag}(t')\widetilde{\rho}(t')\widetilde{s}_{j,i}(t')\}
\nonumber
\\
&&-\sum_{k,i\in 1,2,j\in
L,R}\int_{0}^{t}dt'\thinspace|V_{k}^{j}|^{2}[1-f^{j}(\epsilon_{k}^{j})]e^{-i\epsilon_{k}^{j}(t-t')}\{\widetilde{\rho}(t')\widetilde{s}_{j,i}^{\dag}(t')\widetilde{s}_{j,i}(t)-\widetilde{s}_{j,i}(t)\widetilde{\rho}(t')\widetilde{s}_{j,i}^{\dag}(t')\}
\nonumber
\\
\label{eqcomplete}
\end{eqnarray}
\end{widetext}
where
$f^{j}(\epsilon_{k}^{j})=Tr_{res}\{R_{0}c_{k,j}^{\dag}c_{k,j}\}$ are
the Fermi distributions of each contact ($R_{0}$ is the density
matrix of the electron reservoirs, considered in thermal
equilibrium).\cite{brandesrep,brandes99,platero} Eq.
(\ref{eqcomplete}) can be simplified by rewriting the sums over $k$
as integrals
$\sum_{k}|V_{k}^{j}|^{2}f^{j}(\epsilon_{k}^{j})e^{i\epsilon_{k}^{j}(t-t')}=\int_{-\infty}^{\infty}\frac{d\epsilon}{2\pi}\Gamma_j(\epsilon)f^{j}(\epsilon)e^{i\epsilon(t-t')}$,
%
where $\Gamma_j(\epsilon)\equiv
2\pi\sum_{k}|V_{k}^{j}|^{2}\delta(\epsilon-\epsilon_{k}^{j})$ are
the tunneling rates in and out of the DQD. Working in an "infinite
bias regime" between the reservoirs (such that $f^{L}\rightarrow 1$
and $f^{R}\rightarrow 0$) and assuming a constant density of states
in the reservoirs, the coupling with the leads becomes Markovian:
$\sum_{k}|V_{k}^{L}|^{2}f^{L}(\epsilon_{k}^{L})e^{i\epsilon_{k}^{L}(t-t')}=\Gamma_{L}\delta(t-t')$
and
$\sum_{k}|V_{k}^{R}|^{2}[1-f^{R}(\epsilon_{k}^{R})]e^{i\epsilon_{k}^{R}(t-t')}=\Gamma_{R}\delta(t-t')$
%
and, therefore, Eq. (\ref{eqcomplete}) reads

\begin{widetext}
\begin{eqnarray}
\frac{d}{dt}\widetilde{\rho}(t)=&&-i[\widetilde{H}_T(t),\widetilde{\rho}(t)]\nonumber\\
&&-\frac{\Gamma_L}{2}\sum_{i\in1,2}\{\widetilde{s}_{L,i}(t')\widetilde{s}_{L,i}^{\dag}(t')\widetilde{\rho}(t')-2\widetilde{s}_{L,i}^{\dag}(t')\widetilde{\rho}(t')\widetilde{s}_{L,i}(t')+\widetilde{\rho}(t')\widetilde{s}_{L,i}(t')\widetilde{s}_{L,i}^{\dag}(t')\}
\nonumber
\\
&&-\frac{\Gamma_R}{2}\sum_{i\in
1,2}\{\widetilde{s}_{R,i}^{\dag}(t')\widetilde{s}_{R,i}(t')\widetilde{\rho}(t')-2\widetilde{s}_{R,i}(t')\widetilde{\rho}(t')\widetilde{s}_{R,i}^{\dag}(t')+\widetilde{\rho}(t')\widetilde{s}_{R,i}^{\dag}(t')\widetilde{s}_{R,i}(t')\}
\label{eq12}
\end{eqnarray}
\end{widetext}
As we mentioned already, the fact that the coupling with the
reservoirs becomes Markovian in this limit implies, in particular,
that the renormalization of tunneling due to the bosonic bath
becomes ineffective (for example,
$\overline{s}_{L,1}(t')\overline{s}_{L,1}^{\dag}(t')={s}_{L,1}(t'){s}_{L,1}^{\dag}(t')$).

Invariance under unitary operations implies that the expected value
of any dot operator can be written as $\langle
O(t)\rangle=Tr_{dot}\{Tr_{B}\{\widetilde{\rho}(t)\}Tr_{B}\{\widetilde{O}(t)\}\}=Tr_{dot,B}\{\widetilde{\rho}(t)\widetilde{O}(t)\}$,
where $Tr_{B}$ is the trace over the bath states. In particular, the
expected value of the projector operators over the system states
$Y_{nm}=|n\rangle\langle m|$, can be written as $\langle
Y_{nm}\rangle=Tr_{dot}\{\rho^{S}Y_{nm}\}=\langle
m|\rho^S|n\rangle=Tr_{dot}\{\widetilde{\rho}^{S}\widetilde{Y}_{nm}\}$,
where we have defined the RDM of the DQDs array (system) as
$\rho^S=Tr_{B}\rho$. It is therefore possible to obtain matrix
elements of the \emph{reduced density operator} by just calculating
the expectation value for the suitable $\widetilde{Y}_{nm}(t)$
operators \emph{directly} from the master equation (\ref{eq12}).
Using the notation $\rho_{mn}^S\equiv\langle m|\rho^S|n\rangle$, we
obtain the following set of exact equations:

\begin{widetext}
\begin{eqnarray}
\rho_{nm}^S(t)=&&\rho_{nm}^S(0)-i\int_{0}^{t}Tr_{dot,B}\{\widetilde{\rho}(t')[\widetilde{Y}_{nm}(t),\widetilde{H}_T(t')]\}dt'\nonumber\\
&&-\frac{\Gamma_L}{2}\int_{0}^{t}Tr_{dot,B}\{(\widetilde{s}_L(t')\widetilde{s}_{L}^{\dag}(t')\widetilde{\rho}(t')-2\widetilde{s}_{L}^{\dag}(t')\widetilde{\rho}(t')\widetilde{s}_{L}(t')
+\widetilde{\rho}(t')\widetilde{s}_{L}(t')\widetilde{s}_{L}^{\dag}(t'))\widetilde{Y}_{mn}(t)\}dt'\nonumber\\&&-\frac{\Gamma_R}{2}\int_{0}^{t}
Tr_{dot,B}\{(\widetilde{s}_{R}^{\dag}(t')\widetilde{s}_{R}(t')\widetilde{\rho}(t')-2\widetilde{s}_{R}(t')\widetilde{\rho}(t')\widetilde{s}_{R}^{\dag}(t')+
\widetilde{\rho}(t')\widetilde{s}_{R}^{\dag}(t')\widetilde{s}_{R}(t'))\widetilde{Y}_{mn}(t)\}dt'.
\label{eq13}
\end{eqnarray}
\end{widetext}
%

The full expression of the density matrix is too large to give it
here (its total dimension is $6\times6$) and we show just two
examples for $\rho_{14}^S(t)$ and $\rho_{13}^S(t)$ elements:

\begin{widetext}
\begin{eqnarray}
\dot{\rho}_{13}^S(t)=&&-i\{t_ce^{-i(\overline{\varepsilon}_{1}-\overline{\varepsilon}_{3})(t-t')}(\langle\widetilde{Y}_{33}(t')X_{t}^{\dag}X_{t'}\rangle-\langle\widetilde{Y}_{11}(t')X_{t'}X_{t}^{\dag}\rangle+\langle\widetilde{Y}_{32}(t')^{\dag}X_{t}^{\dag}X_{t'}\rangle-\langle\widetilde{Y}_{41}^{\dag}(t')X_{t'}X_{t}^{\dag}\rangle)
\nonumber\\
\dot{\rho}_{14}^S(t)=&&-i\{t_ce^{-i(\overline{\varepsilon}_{1}-\overline{\varepsilon}_{4})(t-t')}(\langle\widetilde{Y}_{43}^{\dag}(t')X_{t'}X_{t}^{\dag}X_{t}^{\dag}X_{t'}\rangle+\langle\widetilde{Y}_{21}^{\dag}(t')X_{t'}X_{t'}X_{t}^{\dag}X_{t'}\rangle
\nonumber\\&&+\langle\widetilde{Y}_{42}^{\dag}(t')X_{t'}X_{t}^{\dag}X_{t}^{\dag}X_{t'}\rangle-\langle\widetilde{Y}_{31}^{\dag}(t')X_{t'}X_{t'}X_{t}^{\dag}X_{t}^{\dag}\rangle),
\label{eqcoupled}
\end{eqnarray}
\end{widetext}
with
$\overline{\varepsilon}_n|n\rangle=\left(\sum_{i}\frac{1}{2}\Delta\varepsilon_i\sigma_z^i-\frac{1}{4}\kappa
\sum_{i,j}\sigma_z^i\sigma_z^j\right)|n\rangle$.

Note that Eqs. (\ref{eqcoupled}) are not closed. They contain
expectation values involving products of dot and boson operators, as
for example
$\langle\widetilde{Y}_{33}(t')X_{t}^{\dag}X_{t'}\rangle=Tr_{dot,B}\{\widetilde{\rho}(t')\widetilde{Y}_{33}(t')X_{t}^{\dag}X_{t'}\}$,
which need to be decoupled. If one is not interested in the system
back action on the bath, the latter can be assumed to remain at
thermal equilibrium at all times.\cite{brandesrep} Therefore, the
reduced density operator can be approximated as
$\widetilde{\rho}(t')\approx\rho_{B}(0)\otimes
Tr_{B}\widetilde{\rho}(t')$. By using this approximation, we can
decouple higher order correlation functions as
$\langle\widetilde{Y}_{33}(t')X_{t}^{\dag}X_{t'}\rangle
\approx\langle\widetilde{Y}_{33}(t')\rangle\langle
X_{t}^{\dag}X_{t'}\rangle$, etc. This decoupling corresponds to the
so-called Non Interacting Blip approximation in the spin-boson
problem \cite{weiss}.

For an equilibrium boson bath one can write the correlation
functions as\cite{brandesrep} $C(t-t')=\langle
X_tX_{t'}^{\dag}\rangle=e^{-\Phi(t-t')}$ with
$\Phi(\tau)=\int_{0}^{\infty}\frac{J(\omega)}{\omega^2}\{(1-\cos(\omega
\tau))\coth(\beta\omega/2)+i\sin(\omega\tau)\}$. Note that in our
problem we also need $C_2(t-t')=\langle X_t X_t
X_{t'}^{\dag}X_{t'}^{\dag}\rangle=e^{-2\Phi(t-t')}$, which appear
from coherences involving interdot processes like
$|1\rangle\leftrightarrow|4\rangle$ (namely,
$|L_1,L_2\rangle\leftrightarrow|R_1,R_2\rangle$, see Eq.
(\ref{eqcoupled})). In principle, terms involving half phases
$e^{-\frac{1}{2}\Phi(t-t')}$ also appear in tunneling processes to
the reservoirs (like, for example,
$|R_1,L_2\rangle\stackrel{\Gamma_R}\rightarrow|0_1,L_2\rangle$) but,
again, they do not contribute in the Markovian limit.

The resulting set of coupled equations can be given in matrix form
as
\begin{equation}
{\bf\rho}_{S}(t)={\bf \rho}_{S}(0)+\int_{0}^{t}\{{\bf
M}(t-t'){\bf\rho}_{S}(t')+{\bf \Gamma}\}dt' \label{eq14}
\end{equation}
where ${\bf\rho}_{S}$ is a vector containing the different matrix
elements of the reduced density operator, the vector $\bf{\Gamma}$
contains the terms related with the coupling of the first DQD to the
reservoirs, and ${\bf M}(t-t')$ is a non-Markovian time dependent
kernel which contains the bath correlation functions.\cite{ramon04}
Equation (\ref{eq14}) can be solved in the Laplace space
\cite{brandesrep,ramon04} as
\begin{equation}
{\bf\rho}_{S}(z)=[z-z{\bf M}(z)]^{-1}({\bf\rho}_{S}(0)+{\bf
\Gamma}/z)\label{eq14b}.
\end{equation}
The kernel ${\bf M}(z)$ contains the Laplace transform of the bath
correlation functions
$C_{\varepsilon}^{(*)}=\int_{0}^{\infty}e^{-z\tau}e^{(-)i\varepsilon\tau}C^{(*)}(\tau)d\tau$
evaluated at different energies corresponding to the involved
transition. \cite{ramon04,brandesrep,brandes99}

\section{Entanglement} The full time-dependent density matrix can be obtained
by solving algebraically Eq. (\ref{eq14b}) and performing an inverse
Laplace transformation, which is a formidable task. Fortunately, the
entanglement generated by the bath is finite at long times, namely
in the \emph{stationary state}, as we will show. The stationary
solution of Eqs. (\ref{eq14}), $\rho_{\infty}$, is obtained by
extracting the $1/z$ coefficient in a Laurent series of
$\rho_{S}(z)$ for $z\rightarrow0$
.\cite{brandesrep,ramon04,brandes99} For entanglement quantification
we use Wootters' concurrence \cite{wootters98} for a general state
of two qubits,

\begin{equation}
C=max\{0,\sqrt{\lambda_1}-\sqrt{\lambda_2}-\sqrt{\lambda_3}-\sqrt{\lambda_4}\}
\label{eq15}
\end{equation}
where the $\lambda s$ are the eigenvalues in decreasing order of the
non hermitian matrix
$\rho_{S}(\sigma_y\otimes\sigma_y)\rho_{S}^{*}(\sigma_y\otimes\sigma_y)$.
The concurrence ranges from $C=0$ for non-entangled states to $C=1$
for the maximum degree of entanglement. That maximum entanglement is
showed by the Bell states\cite{wootters98}. In the basis of triplet
and singlet states,
$|S_0\rangle=\frac{1}{\sqrt{2}}(|L_{1}R_{2}\rangle-|R_{1}L_{2}\rangle)$,
$|T_0\rangle=\frac{1}{\sqrt{2}}(|L_{1}R_{2}\rangle+|R_{1}L_{2}\rangle)$,
$|T_+\rangle=|L_{1}L_{2}\rangle$ and
$|T_-\rangle=|R_{1}R_{2}\rangle$, the Bell states read:
$|\Psi^{+}\rangle=|T_0\rangle$, $|\Psi^{-}\rangle=|S_0\rangle$,
$|\phi^{+}\rangle =\frac{1}{\sqrt{2}}(|T_-\rangle+|T_+\rangle)$ and
$|\phi^{-}\rangle =\frac{1}{\sqrt{2}}(|T_-\rangle-|T_+\rangle)$.

Importantly, the stationary density matrix in our problem
corresponds to a transport situation and, therefore, a proper
generalization of concurrence to \emph{non-equilibrium} is needed.
Following Ref. \onlinecite{lambert}, we quantify non-equilibrium
entanglement via the concurrence $C$ of the stationary state
$\hat{P}\rho_{\infty}$, where $\hat{P}$ is the projection onto
doubly occupied states including proper normalization. The
projection $\hat{P}$ corresponds to taking the limit $\Gamma_L\to
\infty$ where both qubits are always occupied with one single
electron. For concreteness, we focus on the zero-temperature case.

\begin{figure}
  \begin{center}
    \epsfig{file=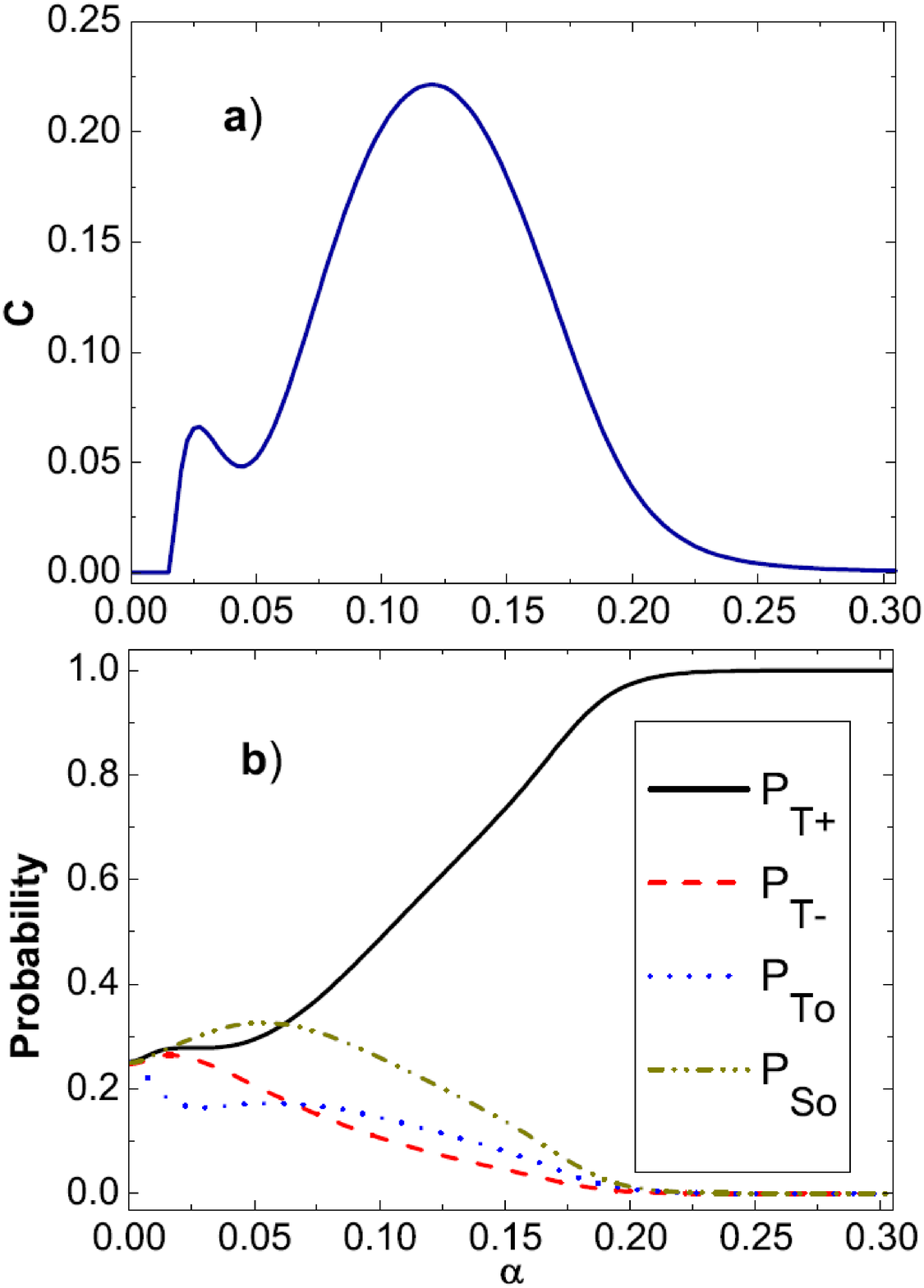,height=9.5cm}
    \caption{(Color online) a) Concurrence as a function of the strength of dissipation
    $\alpha$. b) Population of triplet and singlet states. Parameters: $\Delta\varepsilon_1=\Delta\varepsilon_2=0$, $t_c=3.5$, $\Gamma_L=10$, and
$\omega_c=500$ (in units of $\Gamma_R=1\mu eV$). These parameters
correspond to typical experimental values in  AlGaAs-GaAs lateral
DQDs \cite{vanderWiel,fujisawaapl,fujisawa98}.}
    \label{fig2}
\end{center}
\end{figure}

The concurrence as a function of the coupling $\alpha$ always shows
the same qualitative behavior: for very small $\alpha$ there is a
switching behavior, indicating that below a minimum interaction
strength $\kappa$ the concurrence vanishes, cf. Fig. \ref{fig2}(a)
for identical QDs
$\left(\Delta\varepsilon_1=\Delta\varepsilon_2=0\right)$. As
$\alpha$ increases, two effects compete: entanglement generation and
localization induced by the bath. At small $\alpha$, the two
delocalized states $|S_0\rangle$ and $|T_0\rangle$ have a finite
weight which depends in a nontrivial way on the ratio
$t_c\over\alpha$. On the other hand, for strong coupling the bath
completely freezes the charges on the left dots and the triplet
$|T_+\rangle=|L_{1},L_{2}\rangle$ becomes fully occupied, cf. Fig.
\ref{fig2}(b). These two competing effects lead to the non-monotonic
behavior of the concurrence vs. $\alpha$, with an optimal value at
which the concurrence presents a maximum.

\begin{figure}
  \begin{center}
    \epsfig{file=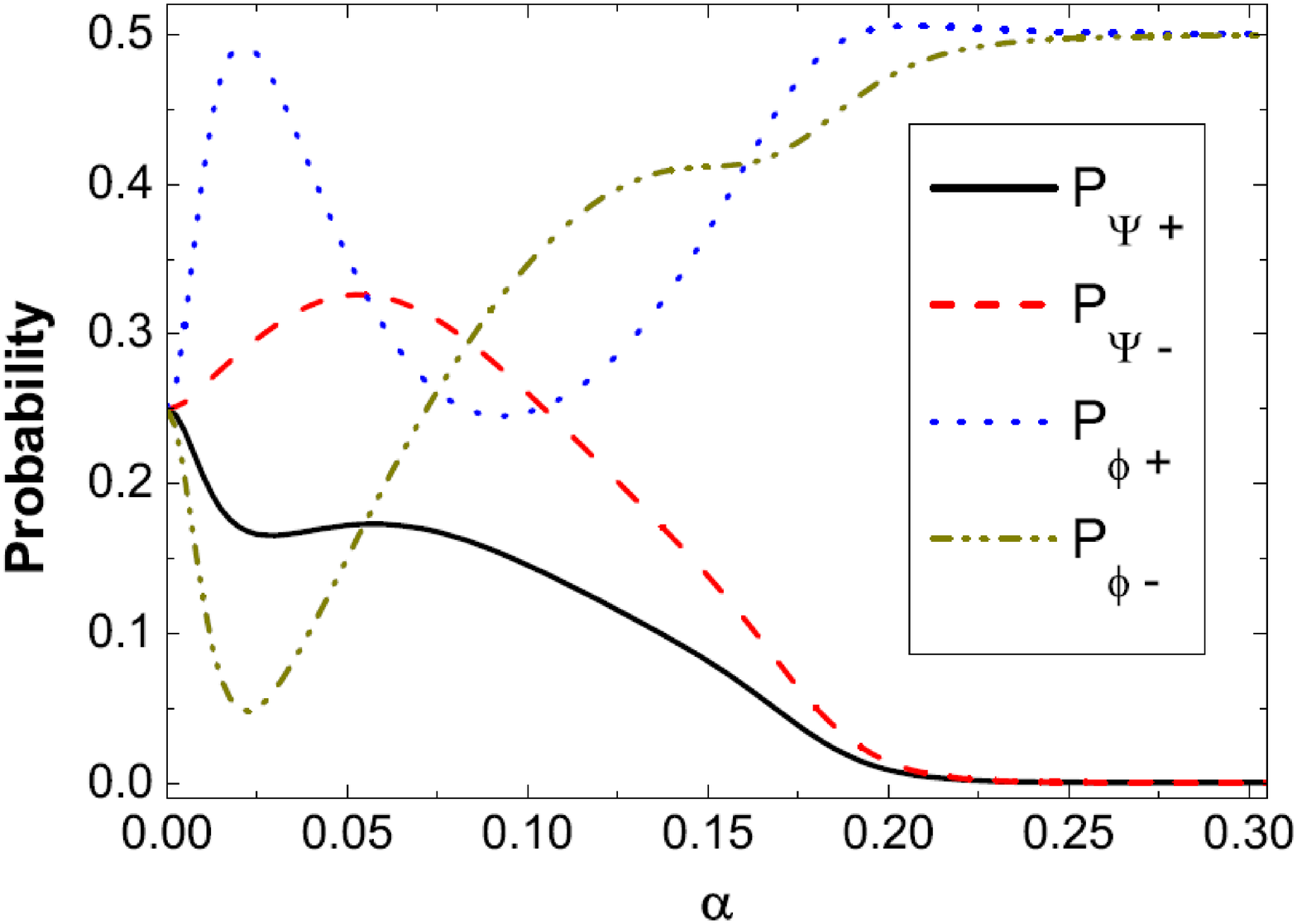,height=6cm}
    \caption{(Color online) Population of the Bell states as a function of $\alpha$. Same parameters as Fig. \ref{fig2}.}
    \label{fig3}
\end{center}
\end{figure}

The population of each Bell state is shown in Fig. \ref{fig3}. The
system does not originate a preferred Bell state and therefore both
maxima in the concurrence contain contributions from all states. The
first concurrence peak is formed by a combination of the four Bell
states with a symmetric contribution of $|\Psi^+\rangle$ and
$|\phi^+\rangle$, whereas on the second peak $|\phi^-\rangle$
probability is slightly dominant. Electrons localization in
"parallel" charge states is reflected in the large probability for
both $|\phi^+\rangle$ and $|\phi^-\rangle$ states for
$\alpha\geq0.2$.

\begin{figure}
  \begin{center}
    \epsfig{file=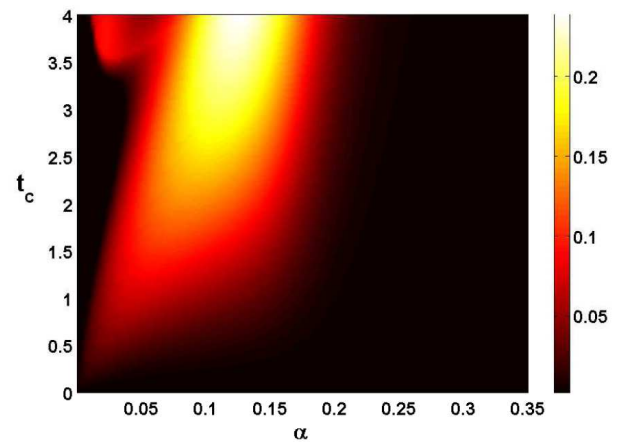,height=6cm}
    \caption{(Color online) Color map of concurrence vs. interdot tunneling $t_c$ and $\alpha$. The rest of parameters are the same as in Fig. \ref{fig2}.}
    \label{fig4}
\end{center}
\end{figure}

The concurrence as a function of both $t_c$ and $\alpha$ is shown in
Fig. \ref{fig4}. For $2t_c<\Gamma_R$, the dephasing induced by the
leads suppresses interdot coherence and the contribution of the
delocalized states $|S_0\rangle$ and $|T_0\rangle$ is negligible.
Thus, the concurrence is almost zero for all $\alpha$. For
$2t_c>\Gamma_R$, entanglement is finite in a region
$\alpha_{min}<\alpha<\alpha_{max}$; both, $\alpha_{min}$ and
$\alpha_{max}$ increase with $t_c$. For $2t_c>>\Gamma_R$, the system
present a maximum in the concurrence at $\alpha\approx 0.15$  with
values $C\approx 0.3$.

\begin{figure}
  \begin{center}
    \epsfig{file=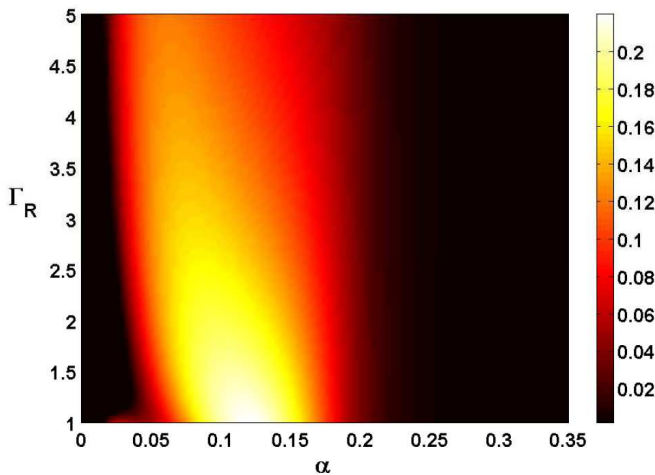,height=6.5cm}
    \caption{(Color online) Color map of concurrence vs. tunneling rate to the right lead $\Gamma_R$ and $\alpha$. The rest of parameters are the same as in Fig. \ref{fig2}.}
    \label{fig5}
\end{center}
\end{figure}
The effect of $\Gamma_R$ on concurrence is shown in Fig. \ref{fig5}.
Here, we also find the switching behavior described above: starting
from $\alpha=0$, the state of the system is strongly mixed for small
$\Gamma_R$. Therefore, $C=0$ below a minimal value $\alpha_{min}$.
This threshold value decreases as $\Gamma_R$ increases. At fixed
$\alpha$, the entanglement decreases as one increases $\Gamma_R$.
For very large $\Gamma_R$, the pure localized triplet
$|T_+\rangle=|L_{1},L_{2}\rangle$ is reached and thus the
entanglement is zero. This effect, which is a transport version of
the Quantum Zeno effect, is similar to the one occurring in
capacitively coupled charge qubits open to reservoirs.\cite{lambert}

\begin{figure}
  \begin{center}
    \epsfig{file=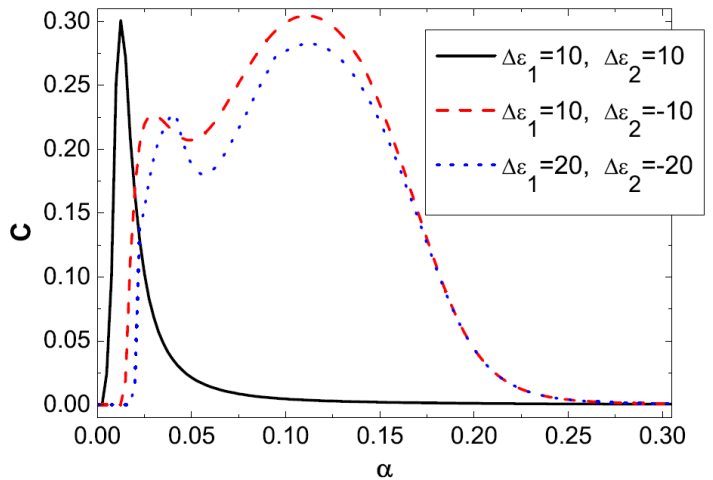,height=6.5cm}
    \caption{(Color online) Concurrence as a function of $\alpha$ for different level detunings. The rest of parameters are the same as in Fig. \ref{fig2}.}
    \label{fig6}
\end{center}
\end{figure}

\begin{figure}
  \begin{center}
    \epsfig{file=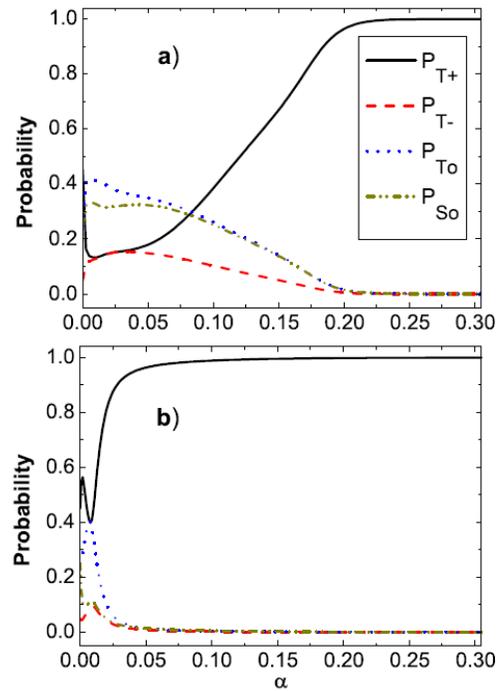,height=9.5cm}
    \caption{(Color online) a) Population of triplet and singlet states for $\Delta\varepsilon_{1}=\Delta\varepsilon_{2}=10$. b) The same for $\Delta\varepsilon_{1}=\Delta\varepsilon_{2}=-10$.}
    \label{fig7}
\end{center}
\end{figure}

A finite detuning $\Delta\varepsilon_{i}>0$
$(\Delta\varepsilon_{i}<0)$ localizes the charge on the lower
(upper) QD of each pair and, therefore, the entanglement should
depend on whether $\Delta\varepsilon_{1}=\Delta\varepsilon_{2}>0$ or
$\Delta\varepsilon_{1}=-\Delta\varepsilon_{2}>0$. The concurrence of
the latter case is very similar to the one for
$\Delta\varepsilon_{1}=\Delta\varepsilon_{2}=0$, and therefore the
population of singlet and triplet states show also the same kind of
behavior, Fig. \ref{fig7}(a). On the contrary, the concurrence for
$\Delta\varepsilon_{1}=\Delta\varepsilon_{2}>0$ is different with a
narrow resonance at small $\alpha$, cf. Fig. \ref{fig6}. This
resonance corresponds to a maximum in the population of the triplet
$T_0$, cf. Fig. \ref{fig7}(b), followed by a fast decay of both
$T_0$ and $S_0$ and an enhanced population of $T_+$ (and, hence,
zero concurrence). The overall qualitative behavior is in agreement
with Ref. \onlinecite{vorrath2003} where the current through two
DQDs coupled to the same phonon bath is analyzed in the Markovian
limit. The indirect interaction due to bath leads to an enhancement
of the inelastic current at
$\Delta\varepsilon_{1}=\Delta\varepsilon_{2}>0$, and a maximum
population of the triplet $T_0$, which is a transport version of the
Dicke effect. In close analogy with the Dicke effect in quantum
optics, this superrradiance should turn into subradiance as the
probability of finding the system in the singlet $S_0$, rather than
in the triplet $T_0$, increases.\cite{vorrath2003} We do not find,
however, the subradiance counterpart in our analysis of concurrence.

\section{CONCLUSION}
We have shown that the strong coupling of two independent qubits
with a common bosonic bath at zero temperature originates
entanglement between the qubits in the \emph{stationary limit}. We
also identify that two effects compete as a function of the coupling
strength with the bath: entanglement generation and charge
localization induced by the bath. These effects lead to a
non-monotonic behavior of the concurrence as a function of the
coupling strength with the phonons.

In addition, the concurrence strongly depends on tunneling and
energy difference on each DQD as well as on the coupling with
external leads, parameters which can be controlled experimentally.

Due to the small concurrence values obtained here $(C<0.5)$, and the
fact that no preferred Bell state is formed, the use of this setup
may not be an optimal choice for entanglement studies in the solid
state realm. Note, however, that this system is the minimal
implementation of a fully tunable two qubit system coupled to a
common bath. From this point of view,  this realization is an
attractive benchmark in which to study the interplay between quantum
coherence, entanglement and decoherence.







\begin{acknowledgments}
We acknowledge F. Rojas and E. Cota for fruitful discussion. DC was
funded by DGAPA-UNAM (project IN114403) and CONACyT ("beca mixta"
funds and project 43673-F). RA acknowledges financial support from
grants MAT2006-03741 (MEC-Spain), 20060I003 (CSIC) and 200650M047
(CAM).

\end{acknowledgments}

\bibliography{references_pol}

\end{document}